\documentclass[aps,prl,reprint,superscriptaddress,showpacs]{revtex4-1}
\usepackage[T1]{fontenc}
\usepackage[latin9]{inputenc}
\usepackage{amsmath}
\usepackage{amssymb}
\usepackage{graphicx}
\usepackage{esint}
\usepackage{bm}
\usepackage{amsfonts,color}
\usepackage{epstopdf} 
\usepackage{hyperref}

\begin{document}

\title{
Strong Proton-Defect Interaction  in Synergistic Irradiation Damage System
}

\author{Ying Zhang}
\affiliation{Microsystem and Terahertz Research Center, China Academy of Engineering Physics,
Chengdu 610200, P.R. China}
\affiliation{Institute of Electronic Engineering, China Academy of Engineering Physics,
Mianyang 621999, P.R. China}

\author{Yang Liu}
\affiliation{Microsystem and Terahertz Research Center, China Academy of Engineering Physics,
Chengdu 610200, P.R. China}
\affiliation{Institute of Electronic Engineering, China Academy of Engineering Physics,
Mianyang 621999, P.R. China}

\author{Jie Zhao}
\affiliation{Microsystem and Terahertz Research Center, China Academy of Engineering Physics,
Chengdu 610200, P.R. China}
\affiliation{Institute of Electronic Engineering, China Academy of Engineering Physics,
Mianyang 621999, P.R. China}

\author{Hang Zhou}
\affiliation{Microsystem and Terahertz Research Center, China Academy of Engineering Physics,
Chengdu 610200, P.R. China}
\affiliation{Institute of Electronic Engineering, China Academy of Engineering Physics,
Mianyang 621999, P.R. China}

\author{Yu Song}
\email{kwungyusung@gmail.com}
\affiliation{Microsystem and Terahertz Research Center, China Academy of Engineering Physics,
Chengdu 610200, P.R. China}
\affiliation{Institute of Electronic Engineering, China Academy of Engineering Physics,
Mianyang 621999, P.R. China}

\author{Su-Huai Wei}
\affiliation{Beijing Computational Science Research Center, Beijing 100193, China}\affiliation{Microsystem and Terahertz Research Center, China Academy of Engineering Physics,
Chengdu 610200, P.R. China}

\begin{abstract}
We experimentally investigated total ionizing dose (TID) response of the input-stage PNP transitor in an operational amplifier LM324N with various initial displacement damage (IDD). We found that, the damage first decreases and then increases with the total dose, which can be exactly described
by a dose-rate-dependent linear generation term, an IDD-dependent linear annealing term, and a dose-rate-dependent
exponential annealing term. 
We demonstrate that, the first two
terms are the well-known TID and injection annealing
effect, respectively,
while the third term stems from the reactions between
the TID-induced protons diffusing from silica and the 
defects generated in silicon,
which implies 
the presence of a unexplored proton-defect interaction between TID and DD effects.
Moreover, we find that such an interaction is as strong as the injection annealing.
Our work show that, beside the well-known Coulomb interaction of trapping charge in silica with carriers 
in silicon, a strong proton-defect interaction 
also plays important role in the synergistic damage. 
\end{abstract}

\date{\today}
\maketitle

\section{Introduction}

Radiating particles induce both ionizing and non-ionizing 
energy depositions in materials.
As a result, the irradiation damage of semiconductor devices contain 
both total ionizing dose (TID) and displacement damage (DD) effects.
For convenience, the damage is usually regarded as an artificial sum of TID and DD,
which can be investigated by gamma and neutron radiation experiments, respectively.
However, recent experiments have demonstrated that, the practical damage is
{either} smaller or bigger than the artificial sum of TID and DD,
which is called as the synergistic effect~\cite{Barnaby2001_IEEETNS48-2074,Barnaby2002_IEEETNS49-2643,
Li2011_IEEETNS57-831,Li2012_IEEETNS59-439,Li2012_IEEETNS59-625,
Li2015_IEEETNS62-1375,Li2015_IEEETNS62-555,Wang2016_NIMPRA831-322}. 
The underlying mechanism is usually attributed 
to the effects of oxide trapping charge ($N_{ot}$) on 
the carrier recombination near the surface 
of the base region. 
In PNP bipolar transistors, the positive  $N_{ot}$ suppresses 
the recombination current by increasing 
the electron density near the base surface and widening the difference of carrier
densities; as a result, a negative synergistic effect arises. 
In NPN devices, a positive synergistic effect arises because
the positive $N_{ot}$ enhances the surface recombination 
current by lowering the hole density near the base surface and reducing the 
{divergence} of carrier densities~\cite{Barnaby2002_IEEETNS49-2643}. 
These mechanisms are consistent with most of the previous observations.

However, contrary experimental results were also obtained 
~\cite{Wang2015_NIMPRA796-108,Wang2016_NIMPRA831-322}. 
When mixed gamma and neutron irradiations are imposed, the observed base 
current degradation of lateral PNP bipolar transistors is found to be
more severe than the artificial sum 
of those damages under individual irradiations. 
This is opposite with the prediction of a negative synergistic effect.
In Ref.~\cite{Li2012_IEEETNS59-439}, depending on the irradiation conditions, 
both reducing and enhancing effects of ionization damage on displacement damage 
were obtained for PNP bipolar transistors.

\begin{figure}[!b]
  \centering
  \includegraphics[width=\linewidth]{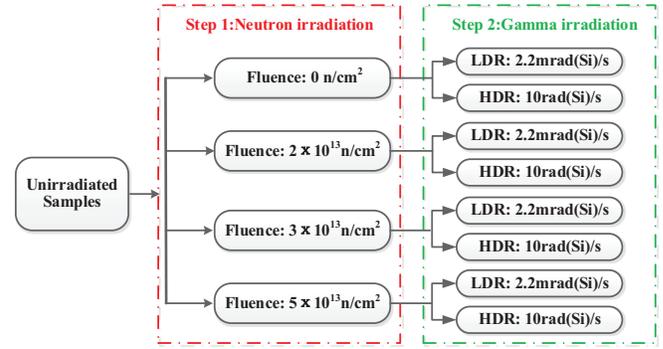}
  \caption{(color online) Flowchart of the neutron/gamma irradiation experiments.
  }\label{fig:expflow}
\end{figure}

\begin{figure}[!b]
  \centering
  \includegraphics[width=0.9\linewidth]{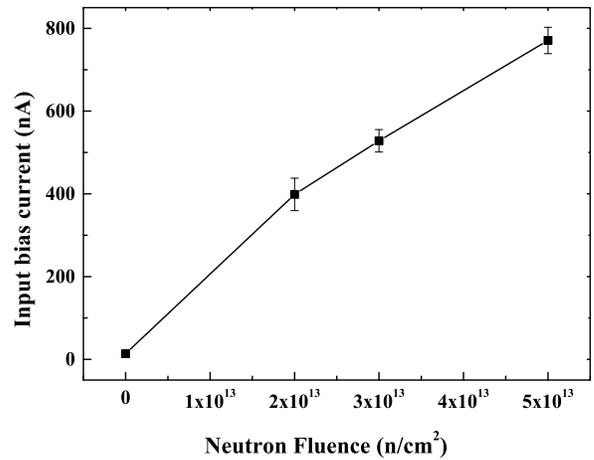}
  \caption{(color online) The input bias current of operational
amplifier LM324N 
  as a function of the neutron fluence. 
  }\label{fig:DDirradiation}
\end{figure}

\begin{figure*}[t!]
    		\centering
    			\includegraphics[width=0.49\textwidth]{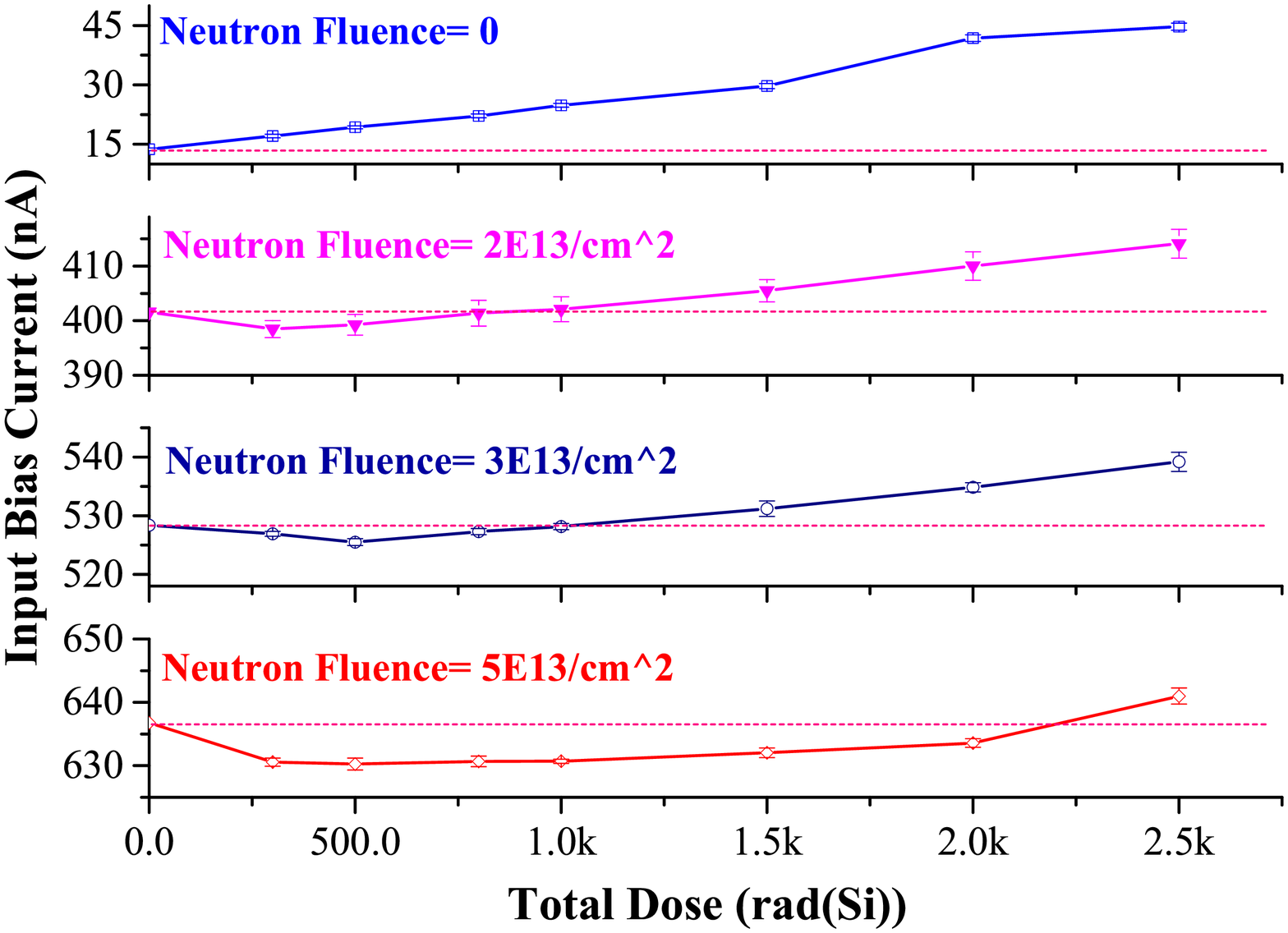}
    			\includegraphics[width=0.49\textwidth]{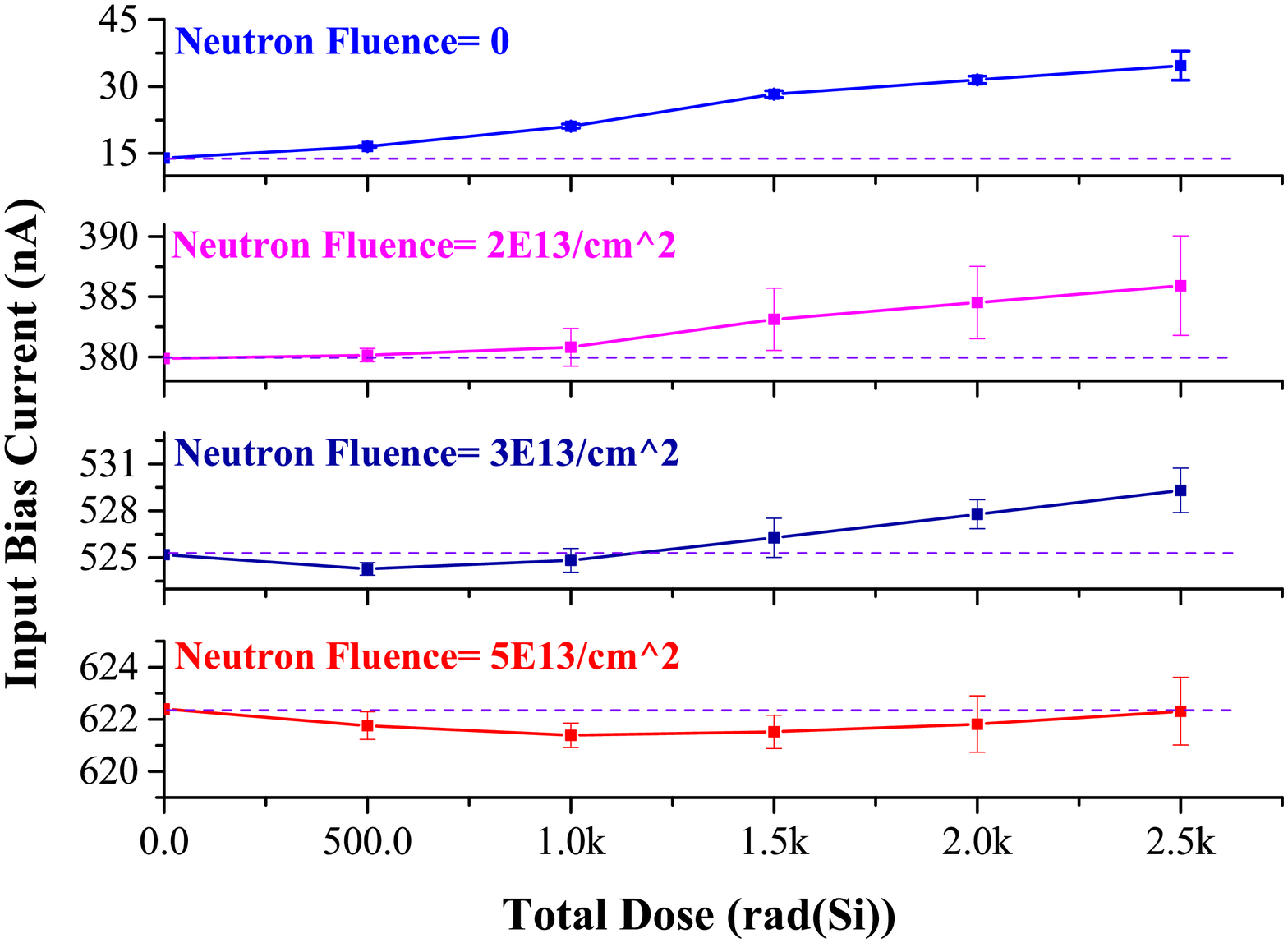}
		\caption{(color online) (a) The input bias current as a function of total dose
  with different IDD for a low dose rate of $2.2~\text{mrad(Si)/s}$.
From top to bottom, IDD reads 
$0$, $2\times 10^{13}~\text{cm}^{-2}$, $3\times 10^{13}~\text{cm}^{-2}$, and $5\times 10^{13}~\text{cm}^{-2}$, respectively.
 (b) The same configurations as (a) but for a high dose rate of $10~\text{rad(Si)/s}$.
} \label{fig:TIDafterDDirradiation}
\end{figure*}

In this work, we show that 
there is another interaction beside the well-known charge-charge interaction. 
We design and carry out experiments on the TID response of 
an operational amplifier 
with various initial DD. 
We find that,
the input bias current first decreases 
and then increases 
with the 
total dose,
which can be exactly described by
a dose-rate-dependent linear generation term, 
an IDD-dependent linear annealing term, and 
a dose-rate-dependent exponential annealing term.
It is found that, the first two terms are exactly
the well-known TID and injection annealing effect~\cite{srour2003review},
while the third term arises because
gamma-induced protons in silica diffuse into silicon and react with few of the
neutron-induced defects. 
This is a direct evidence of the presence of a new interaction between TID and DD effects.
More importantly, 
we find that such a proton-defect interaction can be rather strong,
{compared to the extent} of the injection annealing effect itself.
Our work clearly demonstrates that, beside the well-known charge-charge interaction
between TID and DD effects, 
a proton-defect interaction stemming from diffusion and reaction
may also play {an} important role in the synergistic damage.

\begin{figure*}[t!]
    			\includegraphics[width=0.32\textwidth]{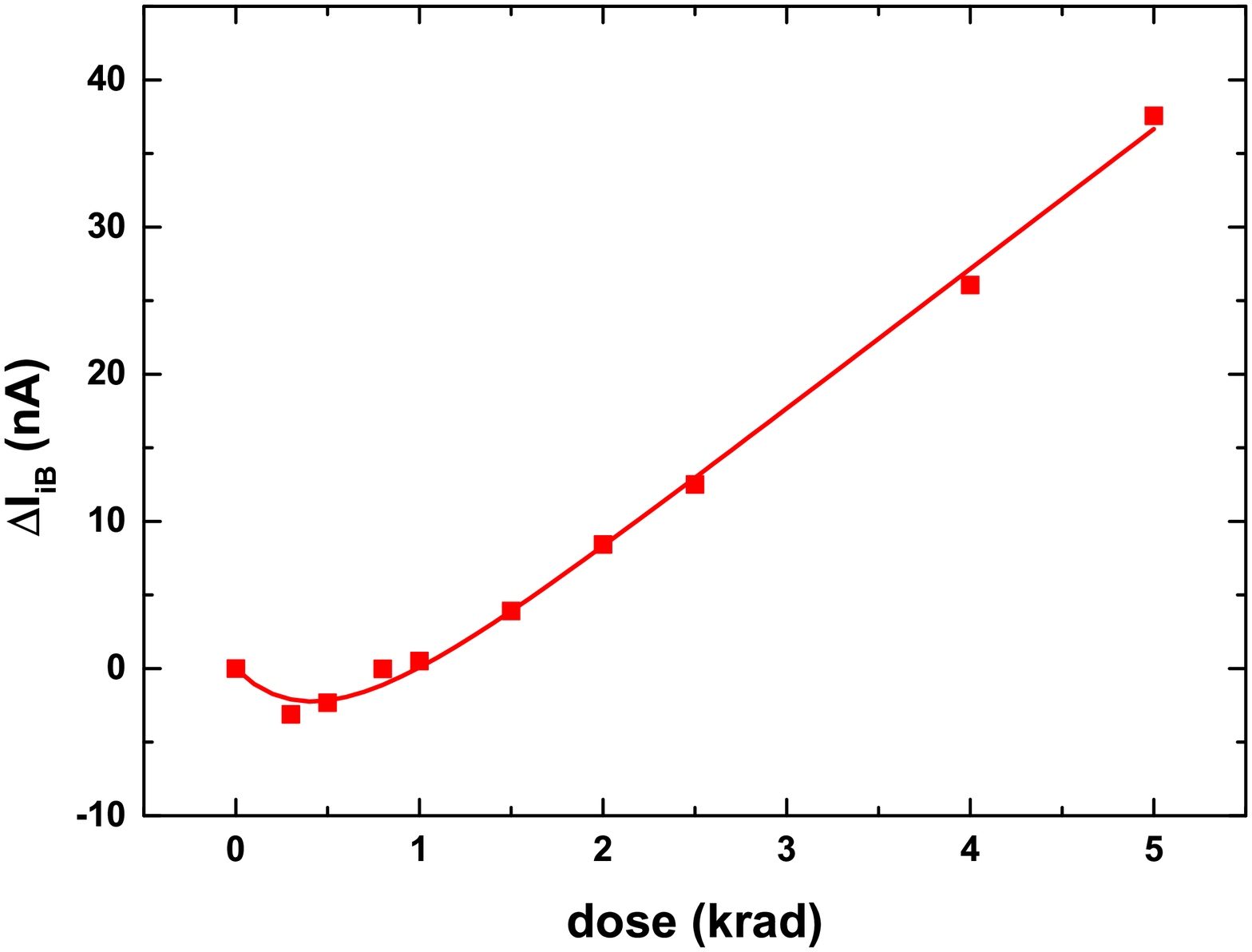}
    			\includegraphics[width=0.32\textwidth]{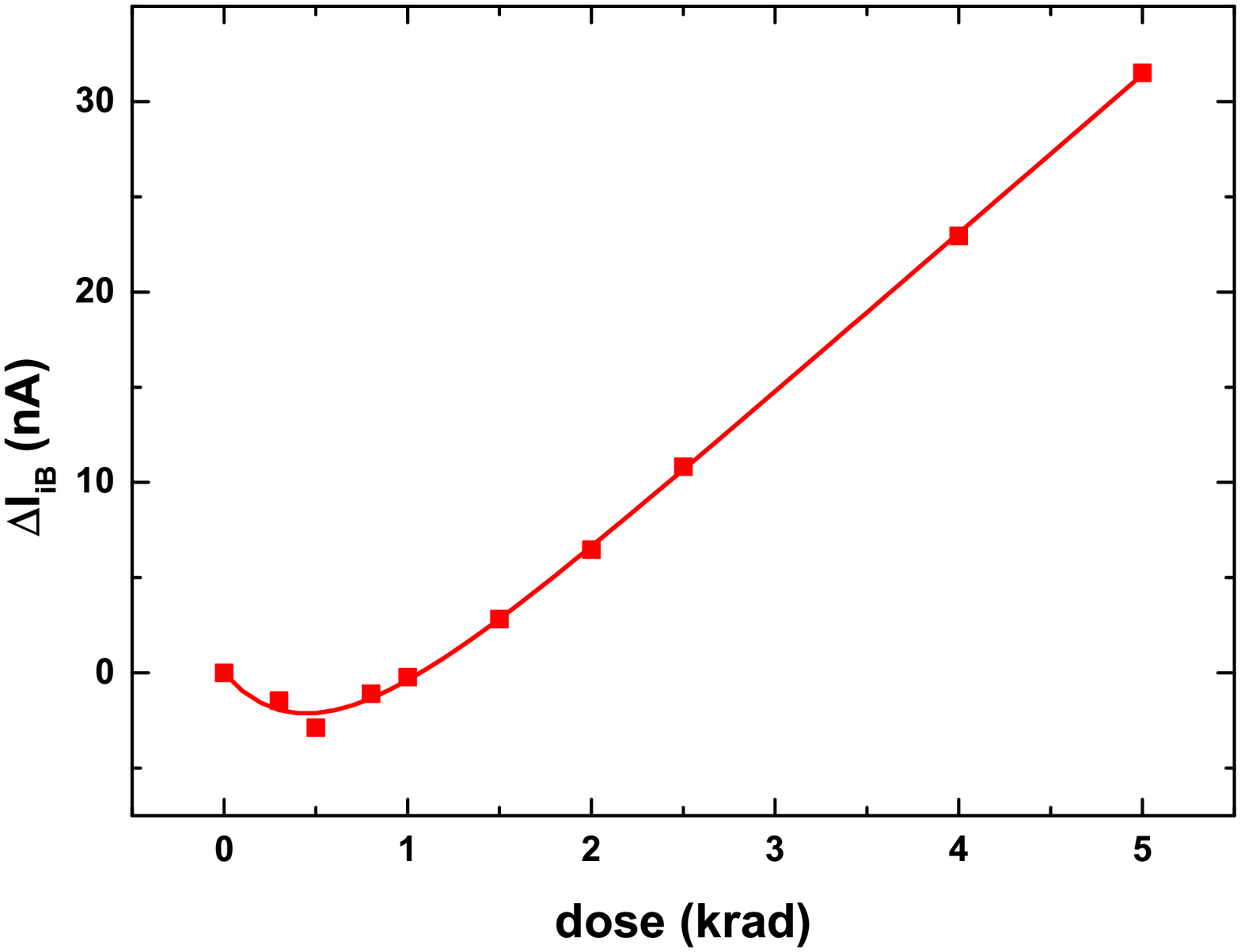}
    			\includegraphics[width=0.32\textwidth]{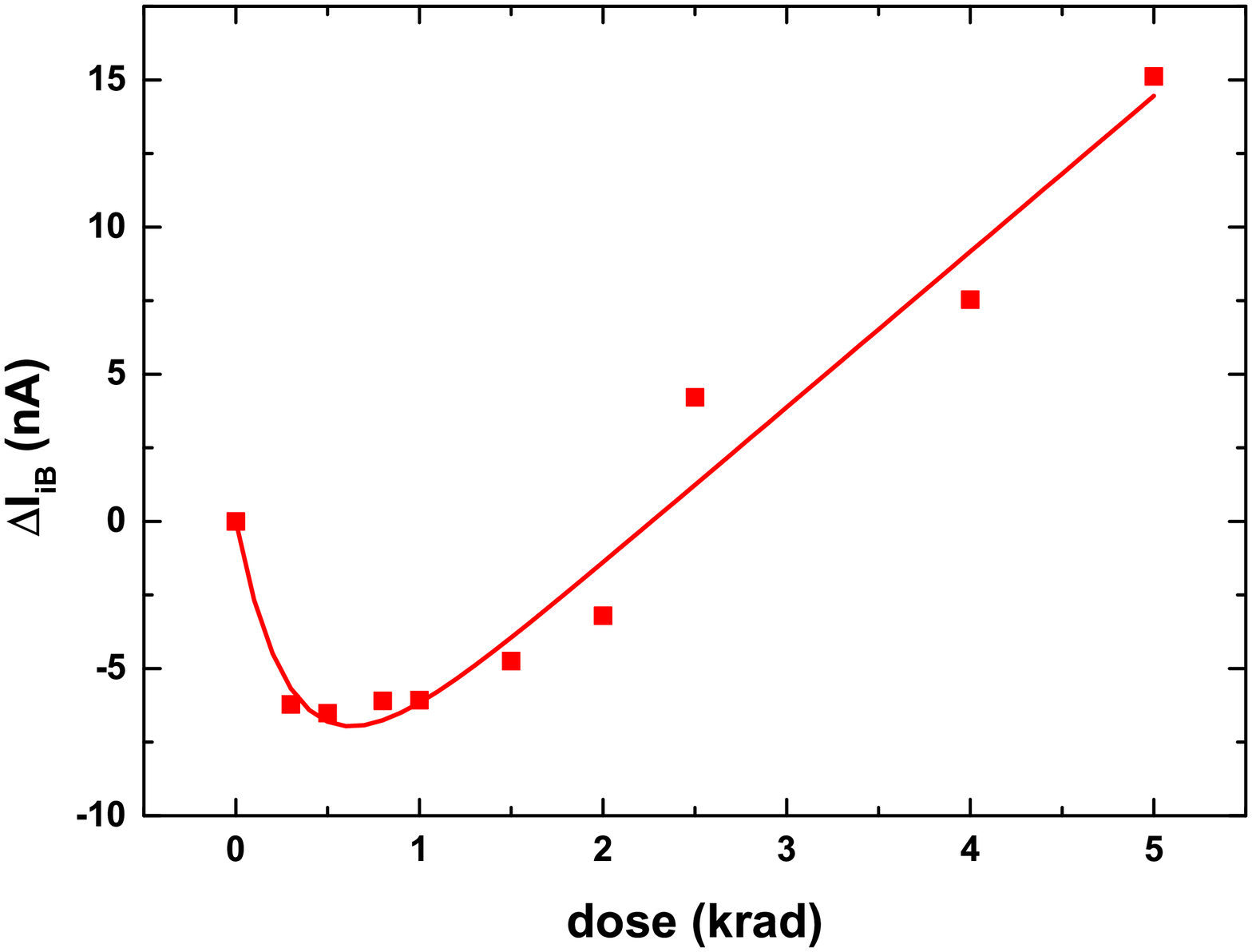}
		~\\
    			\includegraphics[width=0.32\textwidth]{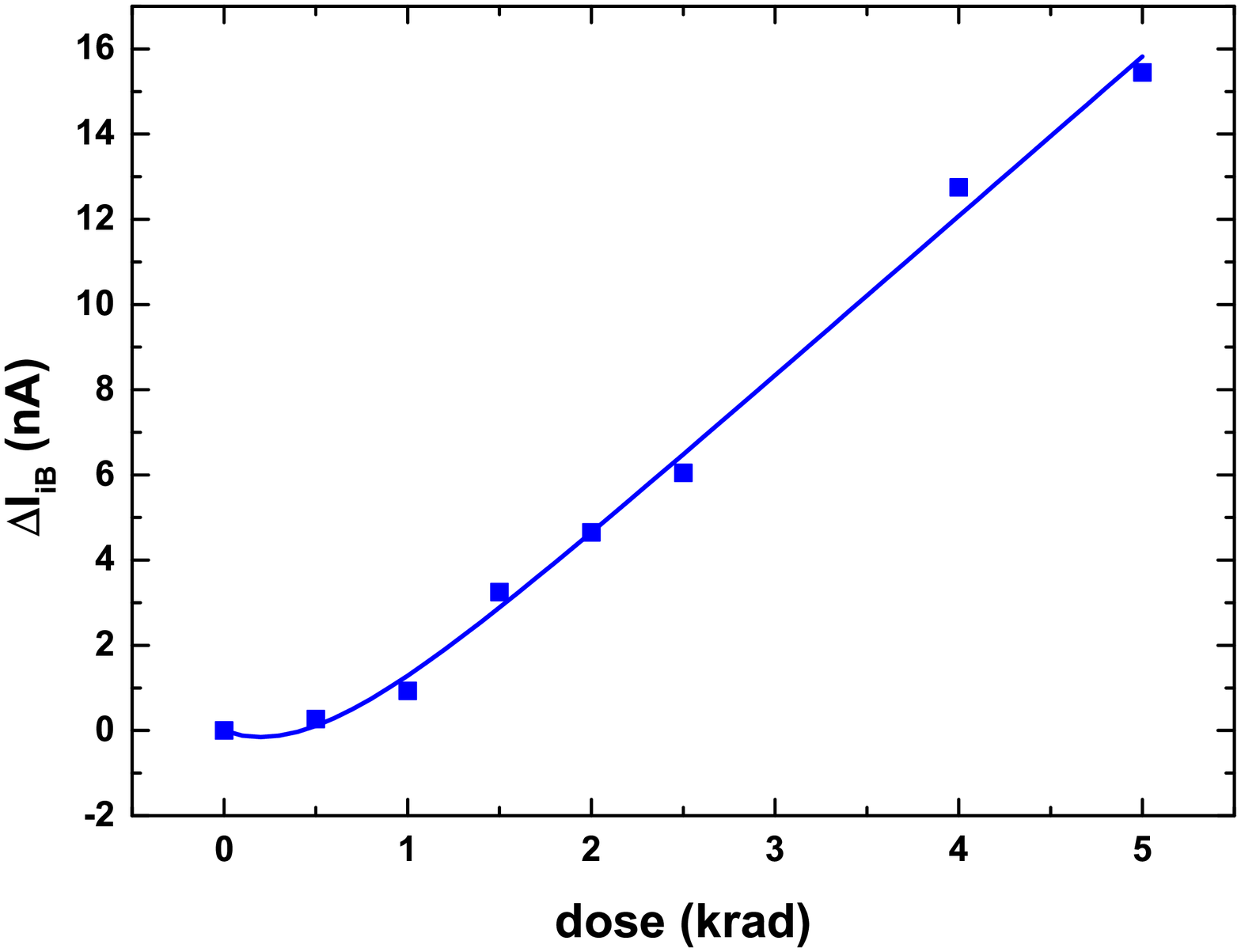}
    			\includegraphics[width=0.32\textwidth]{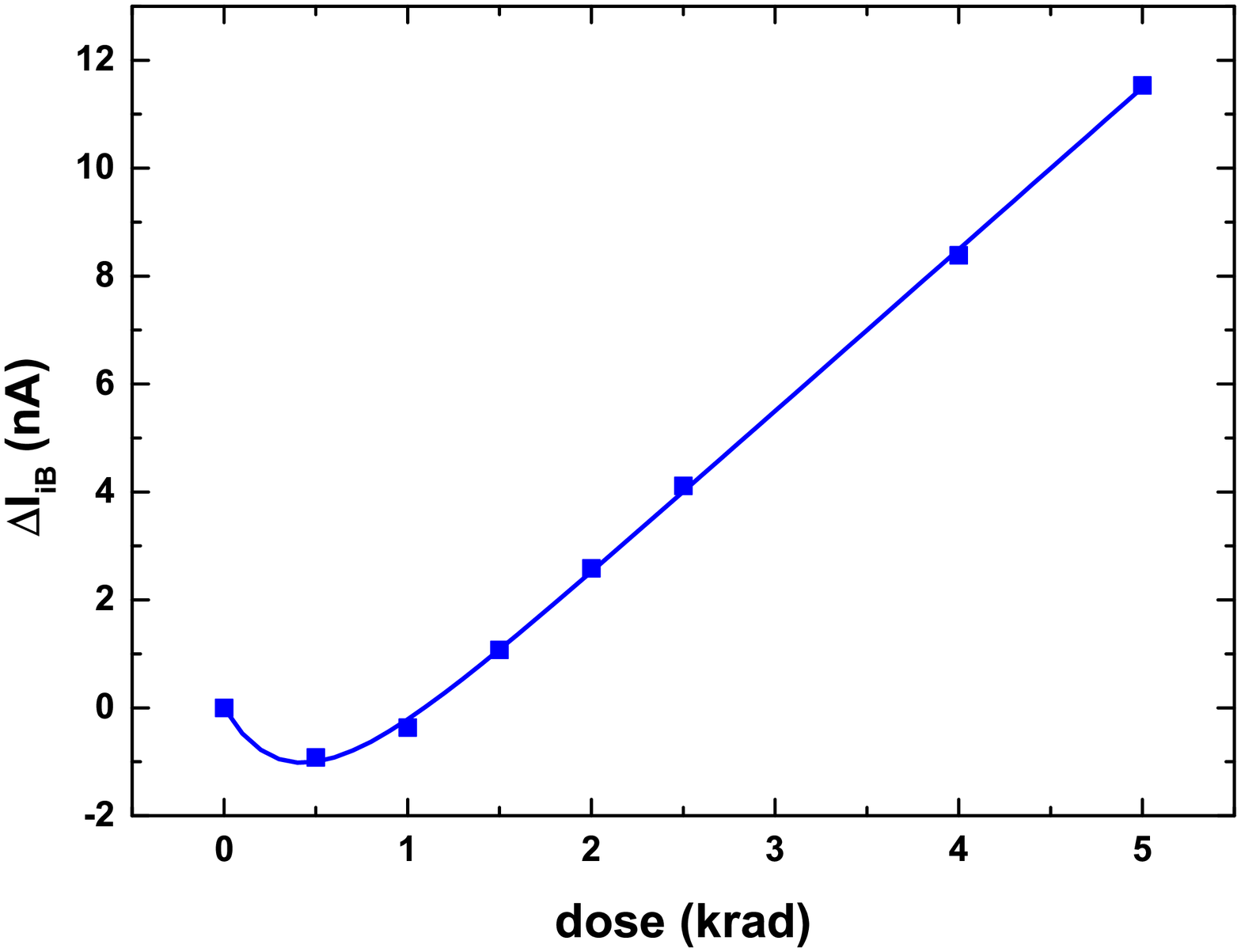}
    			\includegraphics[width=0.32\textwidth]{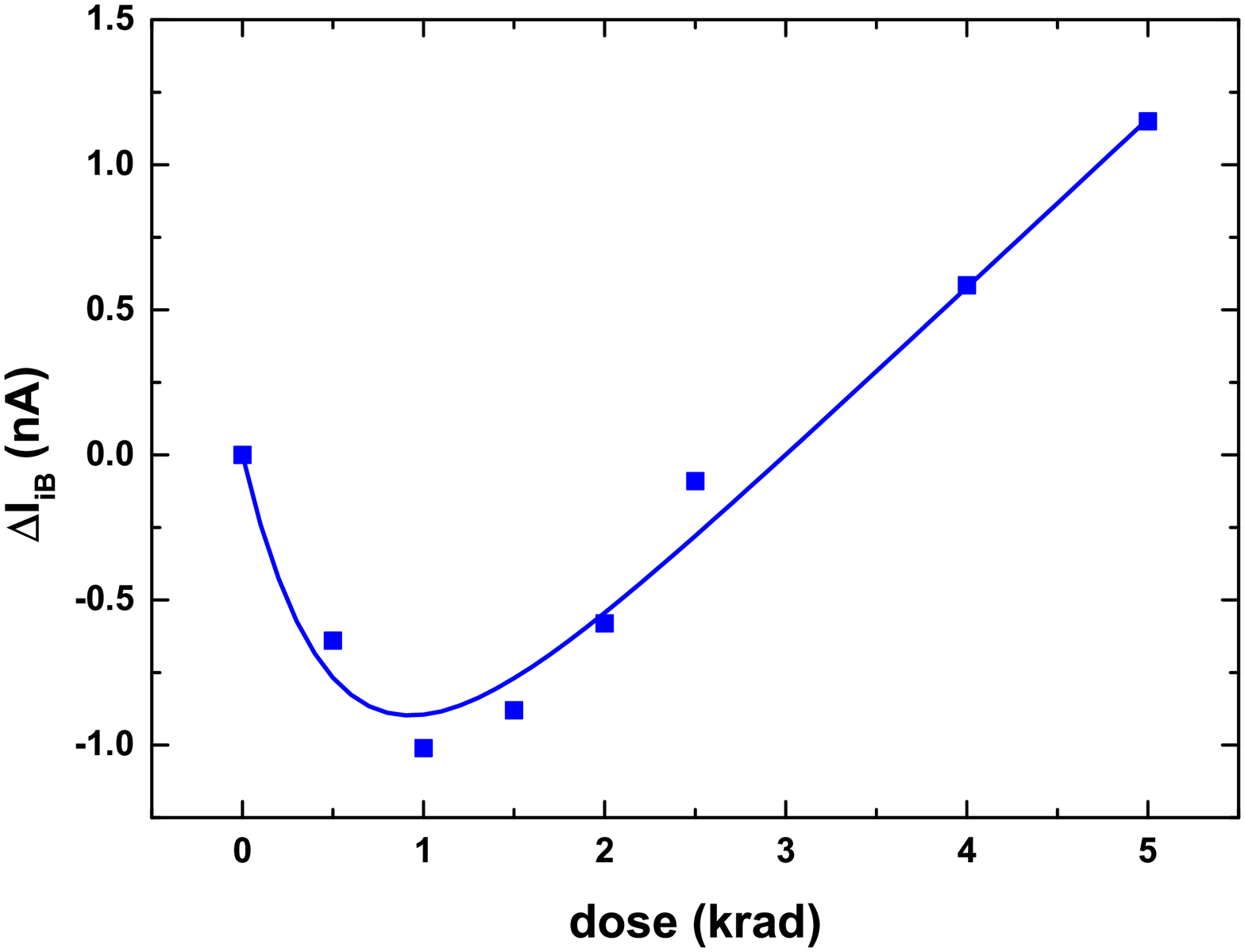}
		\caption{
(color online) The experimental data (dots) and fitting curves (solid) for the input bias current deviation. (a-c) The low dose rate cases of $\gamma$ irradiation with advanced neutron bombardment of the fluence of
$2\times 10^{13}~\text{cm}^{-2}$, $3\times 10^{13}~\text{cm}^{-2}$, $5\times 10^{13}~\text{cm}^{-2}$. (d-f) The high dose rate cases.
} \label{fig:TIDpure}
\end{figure*}

The paper is organized as following.
In Sec. II, we describe the experimental setup.
In the following Sec. III A, we first demonstrate 
the data, which are found can be described by a uniform law
of a linear and an exponential function of the total dose.
We then analysis the origin of the linear term and exponential term in Sec. III B
and Sec. III C, respectively.
The latter is found to {be} an evidence of {the presence} of a non charge-charge interaction
{between TID and DD}.
The relative strength of the proton-defect interaction and key features of the synergistic damage
are investigated in the following Sec. III D and Sec. III E, respectively.
The conculsion is made in Sec. IV.

\section{Experimental setup}\label{sec:model}

To investigate the interaction in synergistic damage, 
an operational amplifier LM324N (Texas Instruments, TI) with a PNP input stage 
was selected for this study. 
{The processes of the experiments are shown in Fig.~\ref{fig:expflow}.}
Neutron irradiations were performed at the Chinese Fast Burst Reactor-II (CFBR-II) of 
Institute of Nuclear Physics and Chemistry, China Academy of Engineering
Physics,
which provides a controlled 1MeV equivalent neutron irradiation. 
Four groups of samples are irradiated 
{with the fluence} of 0, 2$\times$10$^{13}$/cm$^2$, 3$\times$10$^{13}$/cm$^2$, and
5$\times$10$^{13}$/cm$^2$, respectively. 
After that, gamma ray irradiations were done at 
College of Chemistry and Molecular Engineering of Peking University, 
with a high dose rate (HDR) of 10 rad(Si)/s 
and a low dose rate (LDR) of 2.2 mrad(Si)/s, respectively. 
In all the experiments, chips were irradiated in an unbiased configuration with all pins shorted. 
{The changes of the input bias current} 
were measured by BC3193 discrete semiconductor testing systems and used to analyze the damages 
of the input transistors. This is because it
depends directly on the base current of the input transistors.

\section{Results and discussion}

\subsection{Uniform law of the experimental data}

Fig. 2 shows the pure DD response of the devices. 
It is seen that, the input bias current increases sub-linearly 
with the neutron fluence.
When the neutron fluence accumulates from 0 to 5$\times$10$^{13}$/cm$^2$,
the bias current increases from 13.6 nA to 771 nA.
It is well known that, such a degradation is due to a persistent decrease of the lifetime
of minority carrier ($\tau$) with an increase of neutron fluence,
which can be described as \cite{srour2003review}
\begin{equation}
\Delta I_{iB}^{D} = \frac{q n_i A x_{dB}}{2 \tau} e^{\frac{q V_{BE}}{2 k_B T}}~.
\end{equation}
Here the superscript $D$ stands for DD, $q$ is the charge of minority carrier, 
$A$ and $x_{dB}$ are the area and depth of the space charge region, respectively, 
$n_i$ is the concentration of intrinsic carriers,
$V_{BE}$ is the bias between the base and emitter electrodes, 
and $T$ is the temperature.
{Compared with the data as shown in Fig.~\ref{fig:DDirradiation}, we can see that}
$\tau^{-1}$ increases sub-linearly with the neutron fluence.

The low-dose-rate TID response of the devices with various IDD is
shown in Fig.~\ref{fig:TIDafterDDirradiation} (a).
It is seen that, without IDD the TID increases almost linearly with the total dose,
i.e., $\Delta I_B^{I} = k_0 x$, where $x$ is the total dose in unit of krad(Si)
and $k_0 = 14.0$ nA/krad(Si) is obtained from the data.
The relation between  $\Delta N_{it}$ and base current {increment} reads~\cite{schmidt1996modeling} \begin{equation}
\Delta I_{iB}^{I} = \Delta s \frac{q n_i P_E x_{dB}}{2} e^{\frac{q V_{BE}}{2 k_B T}}~,
\end{equation}
where $\Delta s = v_{th} \sigma \Delta N_{it}$ is the surface recombination velocity
induced by interface defects, $v_{th}$ is carrier thermal velocity, $\sigma$ is the carrier capture cross section,
and $P_E$ is the emitter perimeter.
From the data it is seen that $N_{it}$ increases linearly with the total dose.

However, when the IDD becomes nonzero, the damage first decreases 
and then
increases with the total dose. 
For large enough total dose, the damage exceeds the initial one and increases almost linearly. 
In total, the larger the IDD, the critical total dose at which the damage equals to the IDD is larger. 

In following we will focus on the 
divergence of the damage from the IDD after gamma irradiation.
Considering both the linear behavior at high dose in the damage-dose curves
and 
{the exponential-like decline} at low total dose in the dammage-dose curves, 
we {can} use a function containing a positive linear term 
and a negative exponential term to fit the data,
\begin{equation}\label{eq:fitting}
\Delta I_{iB} = a x  - b (1 - e^{- c x}).
\end{equation}
Here $a$ is a factor of the linear term; it has a dimension of nA/krad.
$b$ has a dimension of nA and stands for the strength of the decay term.
$c$ describes the decay rate of the exponential term; 
its dimension is 1/krad.
The fitting curves together with the original data are shown in Fig.~\ref{fig:TIDpure}. 
Surprisingly, although the fitting function is rather simple,
{the data of all six groups are fitted very well.}

\begin{figure*}[t]
    		\centering
    			\includegraphics[width=0.32\textwidth]{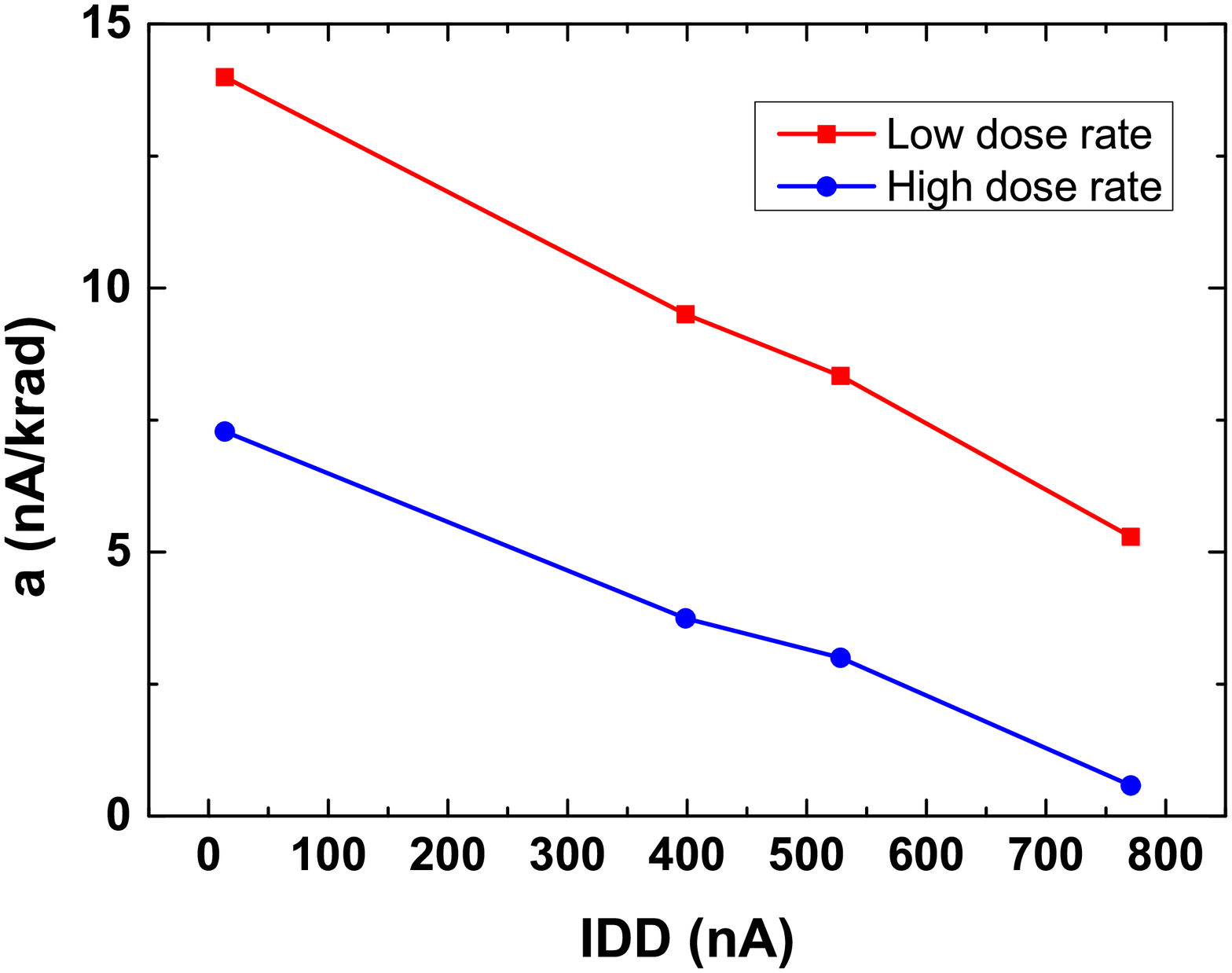}
    			\includegraphics[width=0.32\textwidth]{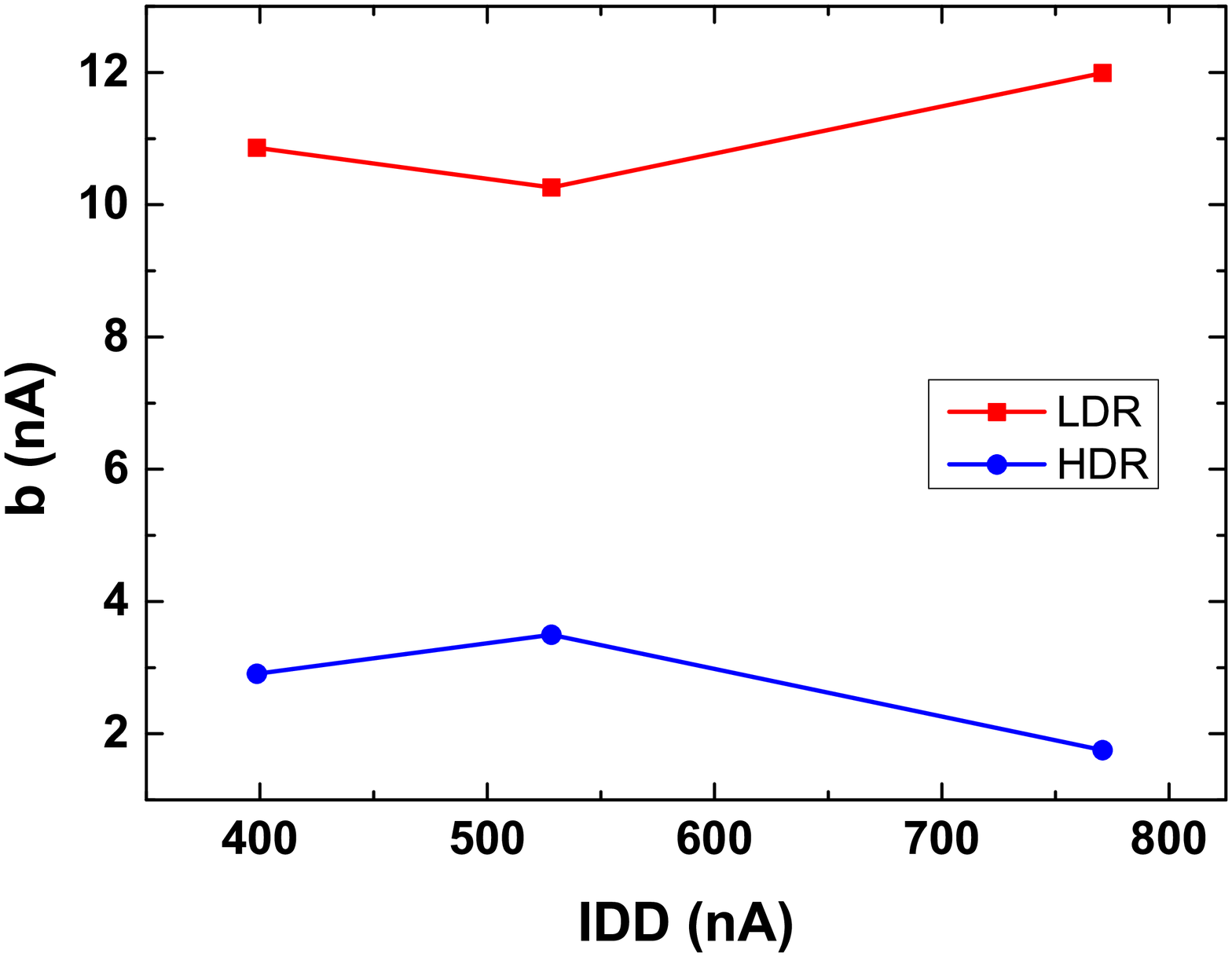}
    			\includegraphics[width=0.32\textwidth]{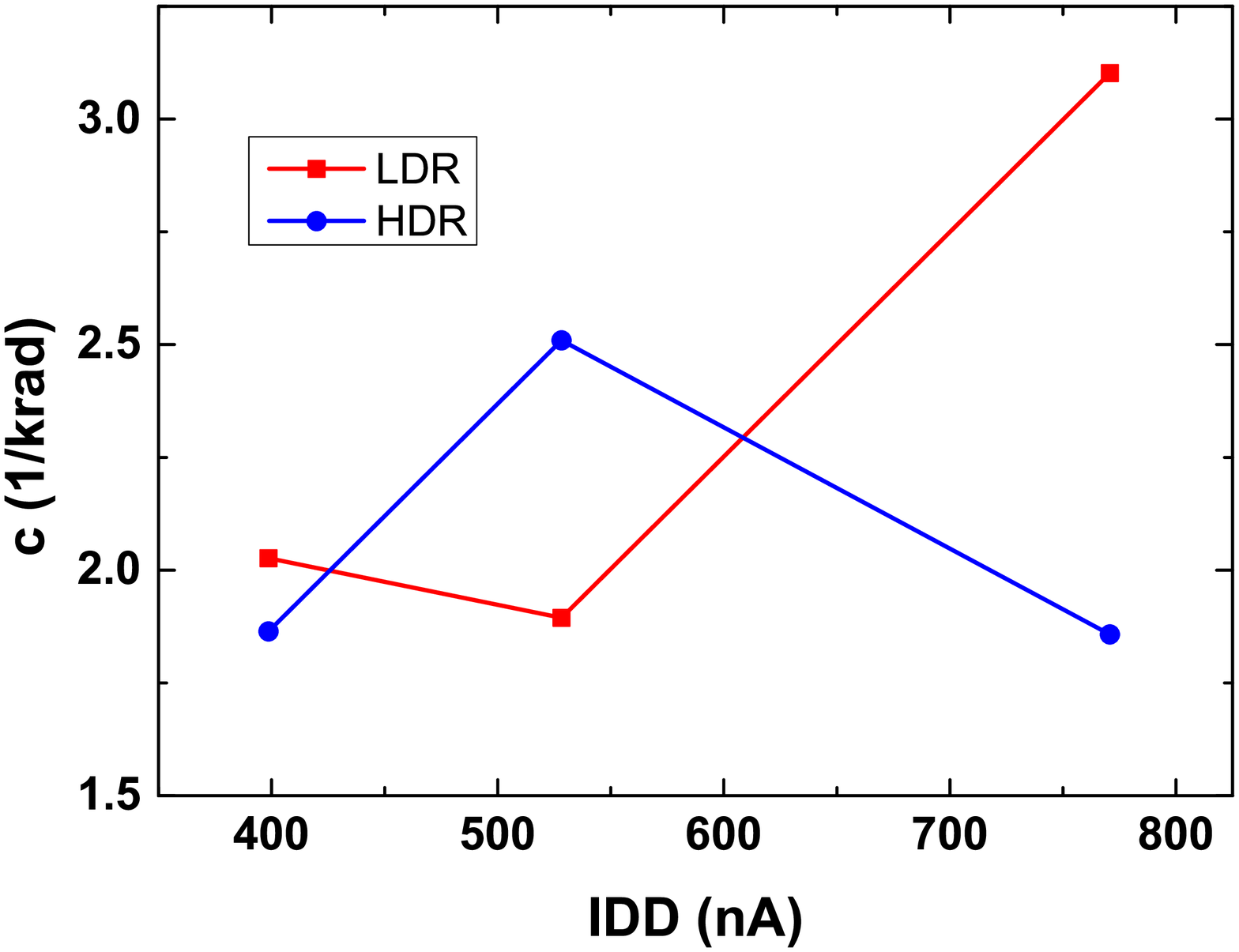}
		\caption{
(color online) 
(a-c) The fitting parameters $a$, $b$, $c$ in Eq.~\ref{eq:fitting} as a function of the inital displacement damage,
which has been converted to the unit of bias current from Fig. 2.
Red for low dose rate and blue for high dose rate.
}\label{fig:linearterm_annealing}
\end{figure*}

\subsection{Origin of the positive linear term
}\label{chap:linearannealing}

Let us first consider what the positive linear term means.
The parameter $a$ as a function of IDD is shown in Fig.~\ref{fig:linearterm_annealing} (a).
With zero IDD, $a$ is equal to $k_0$.
It has larger value for the case of low dose rate irradiation (14.0 nA/krad(Si))
than the case of high dose rate irradiation (7.3 nA/krad(Si)).
This is a representative result of the well-known enhanced 
low dose rate sensitivity (ELDRS) effect.
The enhancement factor is 1.9.
When IDD becomes nonzero, $a$ decreases almost linearly 
with IDD ($D_0$). 
The 
{gradient of the $a$-$D_0$ curves} are almost the same 
for the low and high dose rate cases.
Accordingly, the linear term can be re-written as 
\begin{equation}\label{eq:linearannealing}
\Delta I_{iB}^{lin} = (k_0 - \alpha D_0) x,
\end{equation}
where $\alpha=9.9 \times 10^{-3}/\text{krad}$ and $8.8 \times 10^{-3}/\text{krad}$
are obtained for the low and high dose rate cases, respectively.

{
Now we focus on the second negative term in Eq.~(\ref{eq:linearannealing}).
Because 
in none of the previous experiments have we observed evident damage of neutron irradiation to the silica layers,
the silica layer is thought to be almost transparent to neutron irradiation.
}
Therefore, the second component means that, there is an annealing effect in the system.
In other words, the defects response for damage become fewer.
More importantly, the relation means that the larger the total dose
(the more the initial defects in silicon), 
the faster the damage decreases.
This fact implies 
{that, the annealing is related to} the defects in silicon, 
{where the radiation induced charge carriers play an important role.}
Actually, the enhancement of defects reordering due to injected charge carriers has been
observed in many other experiments~\cite{gregory1967injection,barnes1969thermal,harrity1970short}. 
Mechanisms have been proposed in later studies~\cite{kimerling1975role,kimerling1976new,
bar1984electronic,Bar-Yam1984,car1984microscopic}, 
which claimed that the non-Arrhenius
annealing behaviors result from the enhanced mobility of defects through alternating 
capture and lose electrons.
This mechanism can be expressed as 
\begin{equation}
V_n^- + I^+ =V_{n-1}~,
\end{equation}
{where $V$ is the vacancy and $I$ is the Si interstitial.
Among various defects, the isolated interstitials are mobile particles}.
From this equation, we can see that, 
for a fixed IDD, {the more the charge carriers are excited by the $\gamma$-ray, the more the mobile defects are stimulated,
enhancing the annealing effect.}
On the other hand, for a fixed total dose, the more the defects exist in sample,
the easier to find $I^+$, enhancing the anneal effect as well.
This explains why the term is proportional to both IDD and total dose.
The factor $\alpha$ in Eq. (4) is a reaction velocity between excited vacancy and interstitial.
Comparing the two $\alpha$ values for low and high dose rates,
the enhancement factor for the injection annealing effect is found as 1.125,
which means a rather weak enhance effect.
This is because the role of gamma ray is to make the defects charged, which is a very fast process.

\begin{figure}[!t]
  \centering
 \includegraphics[width=\linewidth]{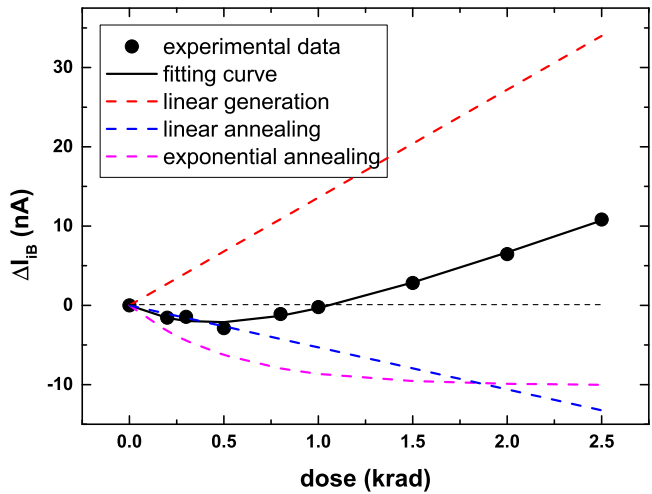}
  \caption{
(color online) The decomposition of the current deviation vs. total dose curve
in Fig. 2(b).
The black solid curve is the fitting to the experimental results while the red, blue, and magenta dashed lines represent the linear generation, linear annealing, and exponential annealing components, respectively.
  }\label{fig:decomposition}
\end{figure}

Now it is clear that, the factor $a$ in Eq. (3) is a sum of a positive linear term 
and a negative linear term,
which are shown as the red and blue dashed lines in Fig.~\ref{fig:decomposition}, respectively.
{The first positive term is the TID effect happening in silica, 
which shows no dependence on IDD.
}
The second negative term is the injection annealing effect of the defects happening in silicon,
which depends on both IDD and total dose in silicon.

\subsection{Origin of the negative exponential term
}

Having understood the positive linear term, we now investigate the origins of
the negative exponential term. 
Such a term is plotted in Fig.~\ref{fig:decomposition} by the magenta dashed curve.
{It can be seen that,} unlike the negative linear injection annealing term,
this term decreases 
with the total dose and saturates at high total dose.
{Compared with Eq.~(\ref{eq:fitting}),} 
the initial velocity is $-b\times c$,
the halflife is $\ln 2/c$, and the maximal value is $-b$.
The parameters as a function of IDD for all six groups of experiments are shown in Fig.~\ref{fig:linearterm_annealing}(b) and (c).
It is interesting that, 
{$b$ shows an evident dependence 
on the dose rate of $\gamma$ irradiation}.
{For lower dose rate, $b$ has a larger value while for higher dose rate, $b$ has a smaller value}.
This phenomena is very similar to the ELDRS effect exists in silica,
in which proton plays central roles.
{Analogously, the strong dose rate dependence of 
the negative exponential component 
suggests the participation of proton
as well.
To test this assumption, we did further analysis in the following.}

Neutron irradiation has been shown to introduce acceptor-like defects in 
Si~\cite{Li1995_IEEETNS42-224,Lutz1995_NIMPRSB95-41,Schulz1994_IEEETNS41-791,
Li1994_NIMPRSA342-105}. 
The divacancies ($V_2$) prominently identified in DLTS measurements~\cite{watts1996new,Lindstrom2001_NIMPRSA466-308}
are thought to be the candidates for these negative charged defect centers~\cite{Li1994_NIMPRSA342-105,Myers2008_JAP104-044507,Watts1996_IEEETNS43-2587}.
Meanwhile, some experiments and theories have shown that,
hydrogen can penetrate the a-SiO2/Si interface~\cite{Pantelides2000_IEEETNS47-2262}
 and diffuse deep into Si with high speed~\cite{Sopori1996_SEMSC41-42-159,Hanoka1986_NSSB136-81,VandeWalle1988_PRL60-2761}.
Further, other experiments and theories also show that,
hydrogen is capable of passivating 
various types of acceptor-like defects and extended defects in Si~\cite{Sopori1996_SEMSC41-42-159,Corbett1991_HS34-49,Hanoka1986_NSSB136-81,
Witczak1998_IEEETNS45-2339,Pearton1985_ICDS13-737,Zhang2001_PRL87-105503,
Johnson1985_PRB31-5525,Assali1985_PRL55-980,Mathiot1989_PRB40-5867,
Pearton1992_HCS,Sana1994_APL64-97,Bourret-Sicotte2017_EP124-267}. 
Many types of complex have been proposed including $V_2H_6$, $V_2H_8$, $VH_2$, $VH$, etc~\cite{Corbett1991_HS34-49,Zhang2001_PRL87-105503,gerasimenko1978infrared}. 
The representative reactions can be described as 
\begin{equation}
V_2 H_n^-  + H^+ \rightleftharpoons V_2 H_{n+1}.
\end{equation}
These reactions
remove the band-gap levels thus reduce the 
recombination rate (decrease the damage).
Estimated from the experimental results,
the popula{tion} of these defects which can be passivated (represented by $D_1\approx$ 5 nA)
is much {less} than the total amount of the defects in silicon (represented by $D_0\approx$ 500 nA),
thus the annealing reaction can cause sensible changes to its concentration.
{In other words, the concentration of the defects is an explicit function of time, $D_1(t)$.}
From Eq. (6), the annealing rate of $D_1(t)$ is $dD_1(t)/dt= - \beta H D_1(t)$, 
where $\beta$ measures the reaction rate.
Integrating the equation, the defect concentration as a function of time
(total dose) is obtained
\begin{equation}
D_1(t)=D_1 e^{-\beta H t}~.
\end{equation}
Accordingly, the bias current has the deviation
\begin{equation}\label{eq:biascurrent}
\Delta I_{iB}^{exp} = - D_1 (1 - e^{- \beta x})~,
\end{equation}
{where we have replaced $H t$ with $x$, which represents the total dose.}
{Eq. (\ref{eq:biascurrent}) has the same form with the exponential term of Eq.~\ref{eq:fitting}.}
Comparing them, we can see that,
the factor $b$ in Eq.~\ref{eq:fitting} is equal to $D_1$, 
which stands for the population of defects that can be passivated by protons;
the factor $c$ in Eq.~(\ref{eq:fitting}) is equal to $\beta$,
which stands for the reaction rate between the defects and protons.

As we have mentioned above,
the values of $b$ (or $D_1$) depends strongly on the dose rate of $\gamma$ irradiation.
{The reasons can be explained similar to ELDRS effect.}
For low dose rate, more protons are generated in silica as a result of ELDRS effect,
so more protons can diffuse into silicon and leads to more reaction.
{
This effect, reflected in Eq.~(\ref{eq:fitting}) and (\ref{eq:biascurrent}) 
as larger values of $b$ (or $D_1$) for lower dose rate.
}
However, the difference of the values of $D_1$ (about 4 times) between the low and high dose rates
is much larger than the enhancement factor of the related ELDRS effect (approximately equal to 2).
The possible reason is that, for low dose rate, there is much longer time for protons
to diffuse, which further increases the amount of the protons in silicon.
Such a large difference supports the diffusion behavior of protons.
{
In contrast to the variable $b$ (or $D_1$),
in Eq.~(\ref{eq:fitting}) and (\ref{eq:biascurrent}),
the variable $c$ (or $\beta$)
shows no evident dependence on $D_0$ nor $D_1$.
}
This is because it measures an intrinsic interaction strength
similar to $\alpha$,
{whose value does not depend on the concentrations of the reactants.}

{Based on the above analysis, we have reasons to believe that,}
the negative exponential term 
stems from the reactions
between protons diffusing from silica and the defects generated in silicon.
This clearly demonstrates that, in the synergistic dammage,
there is also a proton-defect interaction between TID and DD effects.
While the well-known charge-charge interaction leads the decrease (increase) of hole (electron) in silicon, the proton-defect interaction leads to the decrease of defects in silicon.
The strong diffusivity and reducibility of protons play the central role in such an interaction.

\subsection{The relative strength of the proton-defect interaction}

Now an important question would arise as: how strong is the proton-defect  interaction between TID and DD?
Should it be weak enough to the extent that it can be neglected?
We can evaluate this question by making
a comparison with the injection-annealing term.
The injection-annealing reactions happen in the whole silicon bulk,
where the structural defects are introduced by neutron bombardment and 
the charge carriers are excited by $\gamma$ irradiation.
From Eq.~(\ref{eq:linearannealing}),
the strength of the near-linear annealing process is measured by the variable $\alpha D_0$,
whose value is found to be comparable with the damage generation rate of TID effect, measured by $k_0$.
Similarly, the initial strength of the proton-defect interaction is measured by the variable $\beta D_1$.
Seen from Fig.~\ref{fig:decomposition}, the value of $\beta D_1$ is larger than $\alpha D_0$.
This result implies that, the influence of the proton-defect interaction is very strong,
though the population of the defect involved, 
is two order of magnitudes smaller than the total amount of the defects $D_0$.
The reason is that, the reactions in proton-induced annealing
are much easier
than the reactions in charge-induced annealing,
which is reflected in the values of the parameters $\alpha \approx$ 0.01/krad 
and $\beta \approx$ 2/krad.
The underlying mechanism for the strong proton-defect interactions or
fast reactions between protons and the vacancy defects is that,
the $D_1$ defects contain dangling bonds and the binding energy with protons 
are negative~\cite{Corbett1991_HS34-49,Zhang2001_PRL87-105503,gerasimenko1978infrared}.

As the reactions go on, the number of the defects decreases. 
As a result, the strength of proton-defect interaction will decay with the total dose. 
The strength decreases to the half of the initial strength when 
the total dose achieves $x_h = (\ln 2) \beta^{-1}$,
which can be defined as the half-life of the proton-defect interaction 
between TID and DD.
$x_h$ is determined only by the reaction parameter $\beta$. 
For low dose rate, the proton-defect interactions would maintain strong for 
a rather long period of time,
with a half-life in time scale equal to 
\begin{equation}
t_h=\frac{\ln 2 \times \rm {krad}}{2 \times \rm{DR}},
\end{equation}
where DR is dose~rate~in~unit~of~rad/s.
For the case of DR=2.2$\times 10^{-3}$rad(Si)/s, 
the proton-defect interaction is rather strong in a whole day ($t_h \approx$1 day).

\begin{figure}[!t]
  \centering
    			\includegraphics[width=\linewidth]{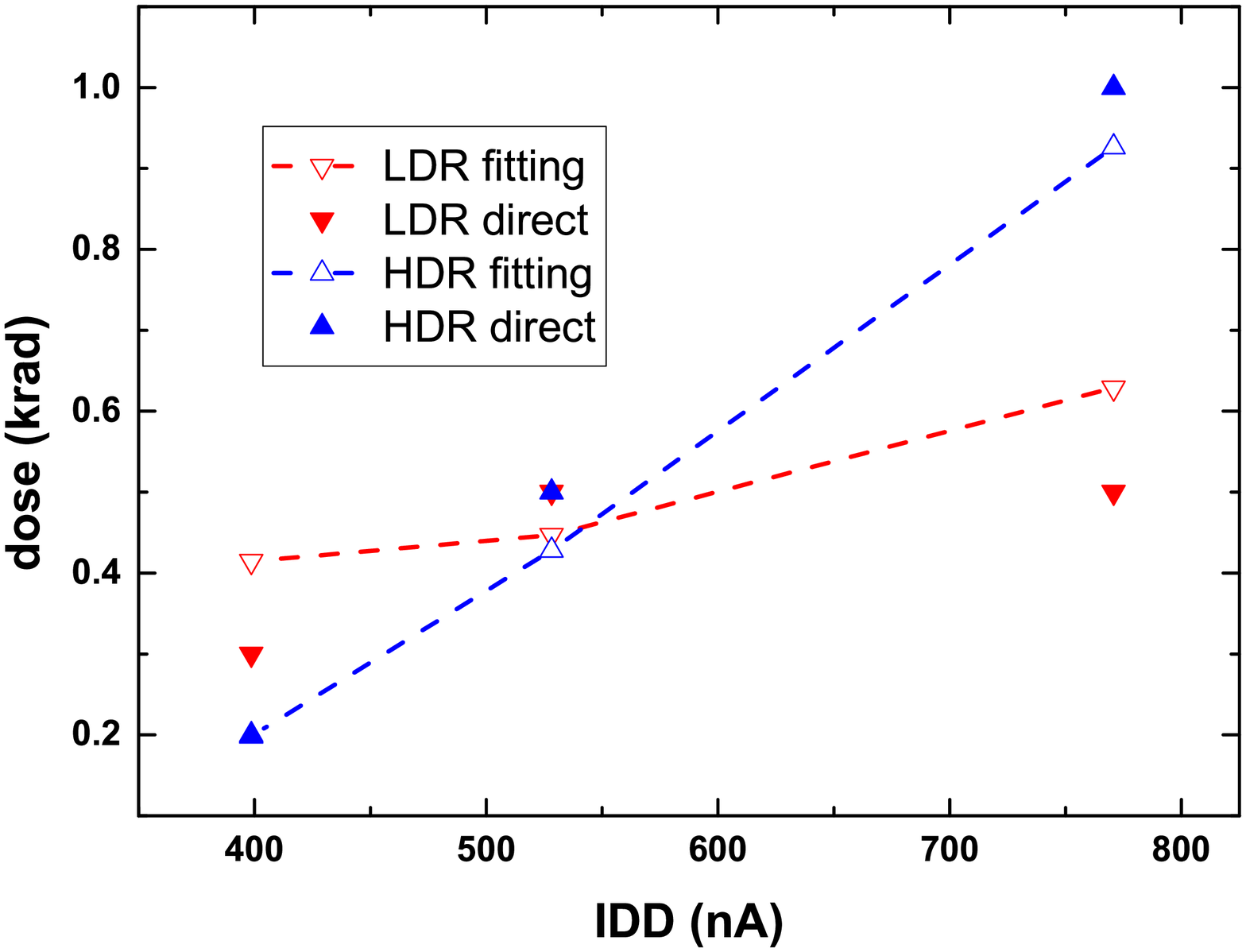}\\
    			\includegraphics[width=0.98\linewidth]{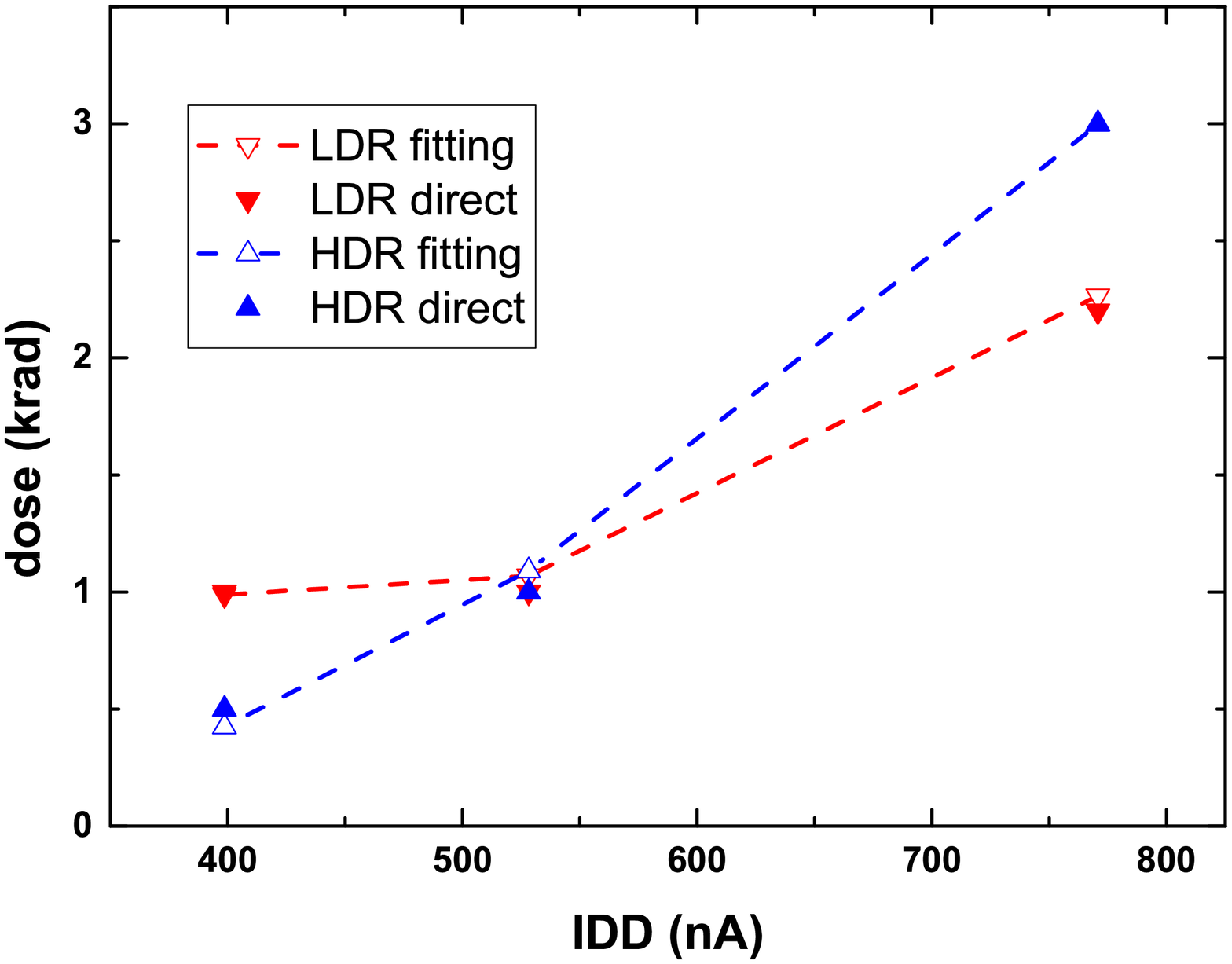}
  \caption{(color online) 
The minimal damage dose (a) and critical dose (b) as a function of total dose.
Dots for experimental data and curves for values calculated by Eq. (12) or (14).
Red for low dose rate and blue for high dose rate.
  }\label{fig:minimaldamage}
\end{figure}

\subsection{Key features of the synergistic damage}

In this last subsection, we will investigate some key features of 
the synergistic damage.
By combining Eqs.~(\ref{eq:fitting}), (\ref{eq:linearannealing}), and (\ref{eq:biascurrent}),
the deviation of the input bias current is described by the following equation:
\begin{equation}\label{eq:deviationcurrent}
\Delta I_{iB}^{syn} = (k_0 - \alpha D_0) x - D_1 (1 - e^{- \beta x})~.
\end{equation}
The first feature we'd like to explain is 
why the deviations of the input bias current (damage) first decrease and then increase,
but not first increases and then decreases, with increasing total dose.
This feature is clearly seen for all curves in Fig.~\ref{fig:TIDafterDDirradiation}.
The changing rates of the input bias currents are derived by
taking derivative of $\Delta I_{iB}^{syn}$ in Eq.~(\ref{eq:deviationcurrent}) with respect to $x$.
The result is
\begin{equation}\label{eq:derivative}
\frac{d \Delta I}{d x} = k \left(1 -  \frac{\kappa}{k} e^{-\beta x}\right)~,
\end{equation}
where $k=k_0 - \alpha D_0$ and $\kappa=\beta D_1$.
We first consider the initial changing rate at $x=0$.
It is readily seen that, 
if $D_1$ is big enough and/or $D_0$ is big enough, 
the second term in the bracket will overwhelm the first term.
In this case, the proton-defect interaction dominates 
and the initial changing rate of the current is negative.
As a result, the damage-dose curves show the declining behaviors.
As the total dose increases, the second term in Eq.~(\ref{eq:derivative}) becomes smaller exponentially. For big enough total dose, the linear term becomes dominates and the changing rate becomes positive.
As a result, the damage-dose curves show the ascending behaviors.
It should be noticed that, if $k$ is too big, 
i.e., $k>\kappa$, 
the initial gradient of 
the damage-dose curves would become positive. 
In this specific limit, there will be no decrease-increase switching behavior. 
With the decrease of the initial neutron fluence,
the value of $D_0$ decreases and the initial decrease of the bias current will become less evident.
The coincident tendency can be seen in Fig.~\ref{fig:TIDafterDDirradiation} by comparing the results of different neutron fluence.

The second feature of the synergistic damage is the presence of a minimal damage.
At a total dose where the derivative of the bias current on the total dose becomes zero, the damage reaches its minimal value.
From Eq.~(\ref{eq:deviationcurrent}) and (\ref{eq:derivative}), 
the total dose is obtained as
\begin{equation}
\beta x_{min}=\ln \left(\frac{\kappa}{k}\right)~,
\end{equation}
and the critical current deviation is
\begin{equation}
\Delta I_{iB}^{min}=D_1 \left[\frac{k}{\kappa}\ln \left(\frac{\kappa}{k}\right)-\left(1-\frac{k}{\kappa}\right)\right]~.
\end{equation}
The calculated values and the experimental results are shown in Fig.~\ref{fig:minimaldamage} (a),
which are found to have good agreements with each other.

The third and last feature is 
that at the critical total dose the input bias current or the equivalent damage
returns to the same amplitude as the initial value before $\gamma$ irradiation. 
Solving Eq.~(\ref{eq:deviationcurrent}) with the condition $\Delta I_{iB}^{syn}=0$, the critical total dose is obtained as
\begin{equation}\label{eq:equaldamage}
\beta x_c = \frac{\kappa}{k}+ W\left[-\frac{\kappa}{k} e^{-\frac{\kappa}{k}}\right]~,
\end{equation}
where $W$ is Lambert-W function or product logarithm.
The calculated values are shown in {Fig.~\ref{fig:minimaldamage}} (b),
which is well coincident with the values directly read from the experimental data.
It can be seen that, for both high and low dose rates,
the critical total dose depends monotonously on the IDD.

It has been noticed that, the ratio of the strength of the proton-defect interactions ($\kappa = \beta D_1$)
and the strength of the linear terms ($k = k_0 - \alpha D_0$) plays crucial roles in all the critical parameters
charactering the current (or damage) behavior of the samples after gamma irradiation.
These results 
reveal the significance of the cooperation of the three mechanisms.

\section{conclusion}

In summary, we have carried out experiments to study the synergistic
TID response of an input-stage PNP transistor in
operational amplifier LM324N with various initial DD. 
We found that, with the increase of the total ionizing dose,
the input bias currents (equivalent to the damages) first decrease and then increase.
Such a behavior can be accurately described by a sum of 
a dose-rate-dependent linear generation term,
an IDD-dependent linear annealing term, and a dose-rate-dependent exponential annealing term. 
The first and second processes are the well-known TID and injection annealing effects,
while the third process, reflected by its rather strong dose rate dependence,
is very likely stems from 
the reactions between protons diffusing from silica and the defects generated in silicon,
hence means the presence of a novel proton-defect interaction
between TID and DD effects.  
The proton-defect interaction is found to be significant in a rather long time period 
limited only by the reaction rate between the defects and protons.
The presence of the proton-defect interaction is also supported by the analysis of key features of the synergistic damage behavior.
The strong diffusivity and reducibility of
protons play the central role in such an interaction.
Our work demonstrates that, 
besides the well-known charge-charge Coulomb interaction, a proton-defect interaction could be also 
important to fully understand the synergistic irradiation response of complex environments.

\section*{Acknowledgements}
The authors would like to thank Professor Chun Zheng of Institute of Nuclear Physics and Chemistry, CAEP for his kindly help in neutron irradiation experiments. This work was supported by the Science Challenge Project under Grant No. TZ2016003-1 and NSFC under Grant No. 11404300. 


\end{document}